\documentclass[prd,twocolumn, showpacs,amsmath,amssymb,reprint,nofootinbib]{revtex4-1}

\usepackage{graphicx}
\usepackage{color}
\usepackage{amsmath}
\usepackage{hyperref}

\definecolor{nicered}{rgb}{0.5,0.,0.}
\definecolor{nicegreen}{rgb}{0.,0.5,0.}
\definecolor{niceblue}{rgb}{0.,0.,0.5}
\hypersetup{colorlinks,citecolor=nicegreen,linkcolor=nicered,urlcolor=niceblue}

\newcommand{\be}{\begin{eqnarray}}
\newcommand{\ee}{\end{eqnarray}}

\begin{document}

\title{Lepton flavor violating \texorpdfstring{$Z'$}{Z'} explanation of the muon anomalous magnetic moment} 

\author{Wolfgang Altmannshofer$^1$, Chien-Yi Chen$^{2,3}$, P. S. Bhupal Dev$^4$, Amarjit Soni$^5$} 
\affiliation{$^1$Department of Physics, University of Cincinnati, Cincinnati, OH 45221, USA}
\affiliation{$^2$Department of Physics and Astronomy,
University of Victoria, Victoria, BC V8P 5C2, Canada} 
\affiliation{$^3$Perimeter Institute for Theoretical Physics, Waterloo, ON N2J 2W9, Canada}
\affiliation{$^4$Max-Planck-Institut f\"{u}r Kernphysik, Saupfercheckweg 1, D-69117 Heidelberg, Germany}
\affiliation{$^5$Physics Department, Brookhaven National Laboratory, Upton, NY 11973, USA}

\begin{abstract}
We discuss a minimal solution to the long-standing $(g-2)_\mu$ anomaly in a simple extension of the Standard Model with an extra $Z'$ vector boson that has only flavor off-diagonal couplings to the second and third generation of leptons, i.e. $\mu, \tau, \nu_\mu, \nu_\tau$ and their antiparticles. A simplified model realization, as well as various collider and low-energy constraints on this model, are discussed. We find that the $(g-2)_\mu$-favored region for a $Z'$ lighter than the tau lepton is totally excluded, while a heavier $Z'$ solution is still allowed. Some testable implications of this scenario in future experiments, such as lepton-flavor universality-violating tau decays at Belle 2, and a new four-lepton signature involving same-sign di-muons and di-taus at HL-LHC and FCC-ee, are pointed out. 
A characteristic resonant absorption feature in the high-energy neutrino spectrum might also be observed by neutrino telescopes like IceCube and KM3NeT.
\end{abstract}

\maketitle

\section{Introduction} \label{sec:intro}

The anomalous magnetic moment of the muon $a_\mu \equiv (g-2)_\mu/2$ is among the most precisely known quantities in the Standard Model (SM), and therefore, provides us with a sensitive probe of new physics beyond the SM (BSM)~\cite{Czarnecki:2001pv, Jegerlehner:2009ry}. 
There is a long-standing $3.6\, \sigma$ discrepancy between the SM prediction~\cite{Hagiwara:2011af, Aoyama:2012wk, Kurz:2016bau} and the measured value of $a_\mu$~\cite{Agashe:2014kda}: 
\begin{equation} \label{eq:gm2}
 \Delta a_\mu \ \equiv \  a_\mu^\text{exp} - a_\mu^\text{SM}  \ \simeq \ (288 \pm 80) \times 10^{-11}\, .
\end{equation}
The uncertainties in the experimental measurement, which come from the E821 experiment at BNL~\cite{Bennett:2006fi}, can be reduced by about a factor of four in the upcoming Muon $g-2$ experiment at Fermilab~\cite{Grange:2015fou}. If comparable progress can be made in reducing the uncertainties of the SM prediction~\cite{Blum:2013xva, Blum:2015gfa, Blum:2015you, Chakraborty:2016mwy}, we will have a definite answer to the question whether or not $\Delta a_\mu$ is evidence for BSM physics. 
Thus from a theoretical point of view, it is worthwhile investigating simple BSM scenarios which can account for the  $(g-2)_\mu$ anomaly, should this endure, and at the same time, have complementary tests in other ongoing and near future experiments. 
With this motivation, we discuss here a simple $Z^\prime$ interpretation of the $(g-2)_\mu$ anomaly. 

A sufficiently muonphilic $Z^\prime$ can address the $(g-2)_\mu$ discrepancy~\cite{Foot:1994vd,  Gninenko:2001hx, Murakami:2001cs, Baek:2001kca, Ma:2001md, Pospelov:2008zw, Heeck:2011wj, Davoudiasl:2012ig, Carone:2013uh, Harigaya:2013twa, Altmannshofer:2014cfa, Tomar:2014rya, Altmannshofer:2014pba, Lee:2014tba, Allanach:2015gkd, Heeck:2016xkh, Patra:2016shz}; however, in order to avoid stringent bounds from the charged lepton sector, while being consistent with a sizable contribution to $(g-2)_\mu$, the $Z^\prime$ coupling must violate lepton universality.\footnote{There are other experimental hints of lepton flavor violation or the breakdown of lepton flavor universality in processes involving muons and taus, e.g. in $B^+\to K^+\ell^+ \ell^-$ decays at the LHCb~\cite{Aaij:2014ora}, in $B \to D^{(*)} \tau \nu$ decays at BaBar~\cite{Lees:2012xj}, Belle~\cite{Huschle:2015rga, Abdesselam:2016cgx} and LHCb~\cite{Aaij:2015yra}, and in the $h\to \mu\tau$ decay at both CMS~\cite{Khachatryan:2015kon} and ATLAS~\cite{
Aad:2015gha} (which however seems to have disappeared in the early run-II LHC data~\cite{CMS:2016qvi, Aad:2016blu}). See e.g. Refs.~\cite{Boucenna:2016wpr, Buttazzo:2016kid, Altmannshofer:2016oaq, Nandi:2016wlp, Bauer:2015knc, Das:2016vkr, Wang:2016rvz, Tobe:2016qhz} for the most recent attempts to explain some of these anomalies. In this work we concentrate 
on $(g-2)_\mu$ and only comment on $h\to \mu\tau$.} 
For instance, a sizable $Z^\prime$ coupling to electrons is strongly constrained over a large range of $Z^\prime$ masses from $e^+e^- \to e^+e^-$ measurements at LEP~\cite{Schael:2013ita}, electroweak precision tests~\cite{Hook:2010tw, Curtin:2014cca}, $e^+e^-\to \gamma \ell^+\ell^-$ (with $\ell=e,\mu$) at BaBar~\cite{Lees:2014xha}, $\pi^0\to \gamma \ell^+\ell^-$ at NA48/2~\cite{Batley:2015lha}, the $g-2$ of the electron~\cite{Pospelov:2008zw}, and neutrino-neutrino scattering in supernova cores~\cite{Manohar:1987ec, Dicus:1988jh}. Similarly, a 
sizable flavor-diagonal $Z'$ coupling to muons is strongly constrained from neutrino trident production $\nu_\mu N\to \nu_\mu N \mu^+\mu^-$~\cite{Altmannshofer:2014pba} using the CCFR data~\cite{Mishra:1991bv}. 
In addition, charged lepton flavor-violating (LFV)  processes, such as $\mu\to e\gamma$,  $\mu\to 3e$, $\tau\to \mu\gamma$, $\tau\to 3e$, $\tau\to ee\mu$, $\tau\to e\mu\mu$, constrain all the lepton-flavor-diagonal couplings of the $Z^\prime$, as well as the flavor off-diagonal couplings to electrons and muons~\cite{Farzan:2015hkd, Cheung:2016exp, Yue:2016mqm, Kim:2016bdu}. There also exist stringent LHC constraints from di-lepton resonance searches: $pp\to Z' \to ee,\mu\mu$~\cite{CMS:2015nhc, Aaboud:2016cth}, $\tau\tau$~\cite{CMS:2016zxk} and $e\mu$~\cite{atlas:emu, CMS:2016dfe}. All these constraints require the flavor-diagonal $Z'$ couplings, as well as the flavor off-diagonal couplings involving electrons to be very small, or equivalently, push the $Z^\prime$ mass scale to above multi-TeV range~\cite{Langacker:2008yv}.

We propose a simplified leptophilic $Z^\prime$ scenario with {\it only} a flavor off-diagonal coupling to the muon and tau sector [see Eq.~\eqref{lagZp} below], which trivially satisfies all the above-mentioned constraints, and moreover, can be justified from symmetry arguments, as discussed below. 
In such a scenario, we find that the most relevant constraints come from leptonic $\tau$ decays in low-energy precision experiments, and to some extent, from the leptonic decays of the SM $W$ boson at the LHC. In particular, we show that the $(g-2)_\mu$ anomaly can be accounted for only with $m_{Z^\prime}>m_\tau-m_\mu$ and by allowing a larger $Z^\prime$ coupling to the right-handed charged-leptons than to the left-handed ones, whereas the lighter $Z'$ scenario (with $m_{Z'}<m_\tau-m_\mu$) is ruled out completely from searches for $\tau\to \mu$+invisible decays. 
We emphasize that the entire allowed range can likely be tested in future low-energy precision measurements of lepton flavor universality in $\tau$ decays at Belle 2, as well as in the leptonic decay of the $W$ boson at the LHC. A striking four-lepton collider signature consisting of like-sign di-muons and like-sign di-taus can be probed at the high luminosity phase of the LHC (HL-LHC) as well as at a future electron-positron collider running at the $Z$ pole. We also point out an interesting possibility for the detection of our flavor-violating $Z^\prime$ scenario by the scattering of ultra-high energy neutrinos off lower-energy neutrinos, which leads to characteristic spectral absorption features that might be observable in large volume neutrino telescopes like IceCube and KM3NeT.

The rest of the paper is organized as follows: in Section~\ref{sec:model}, we present our phenomenological model Lagrangian, which can be justified in a concrete BSM scenario. In Section~\ref{sec:gm2}, we show how the $(g-2)_\mu$ anomaly can be resolved in our LFV $Z'$ scenario. Section~\ref{sec:lfv} discusses the lepton flavor universality violating tau decays for $Z'$ masses larger than the tau mass. Section~\ref{sec:2body} discusses the two-body tau decays for a light $Z'$. In Section~\ref{sec:lhc}, we derive the LHC constraints on our model from leptonic $W$ decays. Section~\ref{sec:lep} derives the LEP constraints from $Z$-decay measurements. Section~\ref{sec:4lepton} presents a sensitivity study for the new collider signature of this model. Section~\ref{sec:icecube} discusses some observational prospects of the $Z'$ effects in neutrino telescopes. Our conclusions are given in Section~\ref{sec:concl}.

\section{A Simplified Model} \label{sec:model}

Our simplified model Lagrangian for the $Z'$ coupling exclusively to the muon and tau sector of the SM is given by  
\begin{eqnarray}
\label{lagZp}
\mathcal L_{Z^\prime} & \ = \ & g_L^\prime \big( \bar \mu \gamma^\alpha P_L \tau +  \bar \nu_\mu \gamma^\alpha P_L \nu_\tau \big) Z^\prime_\alpha \nonumber \\
&& \qquad +\, g_R^\prime \big(\bar \mu \gamma^\alpha P_R \tau\big) Z^\prime_\alpha + {\rm H.c.}\, ,
\end{eqnarray}
where $P_{L,R}=(1\mp\gamma^5)/2$ are the chirality projection operators. 
Due to $SU(2)_L$ invariance, the couplings of the left-handed neutrinos and charged leptons are identical, whereas we do not introduce right-handed neutrinos in order to keep the model minimal.
The left-handed and right-handed couplings $g_L^\prime$ and $g_R^\prime$ could in principle contain $C\!P$ violating phases. We will take into account the complex nature of these couplings in all the equations below; in our numerical analysis however, we will take them to be real for simplicity. We allow different LFV couplings of the $Z'$ to left- and right-handed charged leptons, which will be crucial for the $(g-2)_\mu$ explanation. 

We assume the $Z'$ can acquire mass from the spontaneous breaking of some extra $U(1)'$ symmetry, under which it is charged. The details of the mechanism that generates the $Z^\prime$ mass are irrelevant for our phenomenological purposes, and we treat $m_{Z'}$ as a free parameter in the following. 
Since $U(1)_Y$ is the only flavor-blind $U(1)$ symmetry that is anomaly-free with the SM field content, 
the advantage of the extra $U(1)'$ is that the associated $Z'$ can couple differently to different SM fermion families.

As mentioned above, most of the existing experimental constraints involve first generation fermions, which may be regarded as more `fundamental' in the sense that these comprise ordinary matter around us. Thus, we assume that the couplings of the $Z^\prime$ to the first generation fermions are vanishingly small or non-existent~\cite{Kile:2014jea}, so that all these stringent experimental constraints are readily avoided.\footnote{This can be realized, for instance, in concrete models with a gauged $U(1)_{L_\mu-L_\tau}$ symmetry~\cite{Baek:2001kca, Ma:2001md, Harigaya:2013twa, Altmannshofer:2014cfa, Patra:2016shz}, which is in fact the only anomaly-free $U(1)$ group with nonzero charge assignments to SM neutrinos that can lead to
an experimentally viable light $Z'$ without requiring the addition of any exotic fermions. 
Another possibility is a $U(1)$ group charged under only muon or tau number, but this requires new chiral fermions charged under both $SU(2)_L$ and $U(1)_Y$, as well as under the new $U(1)_\mu$ or $U(1)_\tau$ group.}
If the $Z'$ does not couple universally to quarks, there will be no Glashow-Iliopoulos-Maiani (GIM) suppression of the flavor changing neutral current (FCNC) processes in the quark sector, and the current experimental bounds on neutral meson mixing, such as $K-\bar{K},\, D_0-\bar{D}_0, B_d-\bar{B}_d,\, B_s-\bar{B}_s$~\cite{Isidori:2014rba, Altmannshofer:2014cfa}, as well as FCNC decays of the top, bottom and strange quarks~\cite{Altmannshofer:2014cfa, Fuyuto:2015gmk} will force the $Z'$ couplings to be rather small. 
Therefore, we will assume that the $Z'$ in our case is leptophilic, and more specifically, couples only to second and third generation leptons.  
The phenomenological Lagrangian in Eq.~\eqref{lagZp} can then be justified by imposing an exact discrete symmetry under which~\cite{Foot:1994vd}
\begin{align}
L_\mu \ & \leftrightarrow \ L_\tau \, , \qquad \mu_R \  \leftrightarrow \ \tau_R \, , \nonumber \\
B^\alpha \ & \leftrightarrow \ B^\alpha \, ,  \qquad Z'^\alpha \  \leftrightarrow \ -Z'^\alpha \, ,
\label{discrete}
\end{align}
where $L_\ell \equiv (\nu_\ell, \, \ell)_L$ and $\ell_R$ are respectively the usual $SU(2)_L$ lepton doublets and  singlets in the SM in the gauge eigenstate basis and $B^\alpha$ is the $U(1)_Y$ gauge field.\footnote{The discrete charge assignment in Eq.~\eqref{discrete} would require an extended Higgs sector to give masses to all the charged leptons~\cite{Foot:1994vd}, but this does not affect the $Z'$ phenomenology discussed here.} 
Since the $B^\alpha$ gauge field, and hence, the photon and $Z$ fields are even under the discrete symmetry, we can forbid kinetic $Z-Z'$ mixing and $\gamma - Z^\prime$ mixing to all orders, thus removing a few more stringent experimental constraints, e.g.~from neutrino-electron scattering~\cite{Laha:2013xua} and beam dump experiments~\cite{Alekhin:2015byh}.

\section{Muon Anomalous Magnetic Moment} \label{sec:gm2}

\begin{figure}[t]
\centering
\includegraphics[width=4cm]{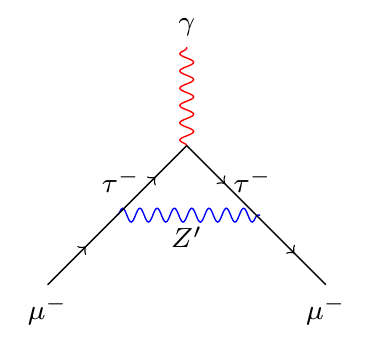}
\caption{Feynman diagram for the $Z'$ contribution to the anomalous magnetic moment of the muon in our model.}
\label{fig:1}
\end{figure}

The flavor-violating $Z^\prime$ coupling in Eq.~\eqref{lagZp} gives rise to a new contribution to $(g-2)_\mu$, as shown in Fig.~\ref{fig:1}, and is given by the general expression~\cite{Leveille:1977rc}\footnote{A typo in Ref.~\cite{Leveille:1977rc} is corrected in the second line of Eq.~\eqref{gm2}.}
\begin{align}
a_\mu \ & = \  \frac{m_\mu^2}{4\pi^2}\int_0^1 dx\Bigg[C_V^2\bigg\{(x-x^2)\left(x+\frac{2m_\tau}{m_\mu}-2\right)\nonumber \\
& \quad  -\frac{x^2}{2m_{Z'}^2}(m_\tau-m_\mu)^2\left(x-\frac{m_\tau}{m_\mu}-1\right)\bigg\}\nonumber \\
& \quad  +C_A^2\bigg \{m_\tau\to -m_\tau\bigg \}\Bigg]\nonumber \\
& \times \Big[m_\mu^2 x^2+m_{Z'}^2(1-x)+x(m_\tau^2-m_\mu^2)\Big]^{-1} \, ,
\label{gm2}
\end{align}
where $C_V=|g'_R+g_L'|/2$ and $C_A=|g'_R-g_L'|/2$ in the notation of Eq.~\eqref{lagZp}. For $m_{Z^\prime} \gg m_\tau$, this reduces to 
\begin{equation}
 a_\mu \ \simeq \ \frac{1}{12\pi^2} \frac{m_\mu^2}{m_{Z^\prime}^2} \left[ 3\: \text{Re}(g_L^\prime g_R^{\prime *}) \frac{m_\tau}{m_\mu} - |g_L^\prime|^2 - |g_R^\prime|^2 \right] \, ,
\label{eq:g-2}
\end{equation}
Note that in the presence of both left-handed and right-handed couplings, the contributions of the flavor changing $Z^\prime$ are enhanced by a factor $m_\tau / m_\mu$. This is in contrast to contributions from flavor-blind new physics, that do not enjoy such an enhancement. Moreover, a  purely left-handed or right-handed coupling would lead to a negative contribution to $a_\mu$, thus making the $\Delta a_\mu$ discrepancy worse than in the SM.   
\begin{figure}[t]
\centering
\includegraphics[width=0.46\textwidth]{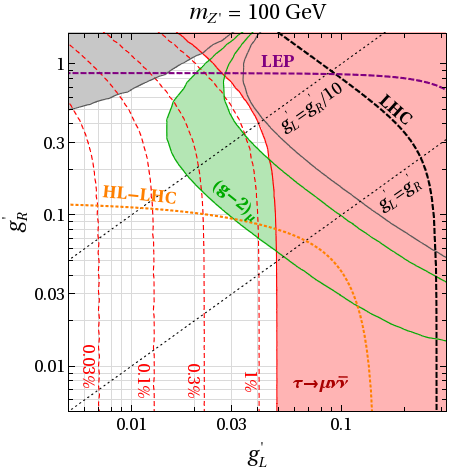}
\caption{The $g_L^\prime$ vs. $g^\prime_R$ plane for $m_{Z^\prime} = 100$~GeV. The green band is preferred at $2\, \sigma$ by the $(g-2)_\mu$ anomaly, whereas the gray region is disfavored at $>5 \, \sigma$ (see Section~\ref{sec:gm2}). The red region is excluded by lepton flavor universality in tau decays (see Section~\ref{sec:lfv}). The dashed red contours show values of constant lepton flavor universality violation in tau decays. The black dashed curve shows the 95\% CL LHC exclusion from searches for leptonic $W$ decays (see Section~\ref{sec:lhc}) and the purple dashed curve shows the 95\% CL LEP exclusion from $Z$ coupling measurements (see Section~\ref{sec:lep}). The orange dotted curve shows the expected $3\sigma$ sensitivity to the process $pp \to \mu^\pm \mu^\pm \tau^\mp \tau^\mp$ at the high-luminosity LHC (see Section~\ref{sec:4lepton}).}
\label{fig:2}
\end{figure}
\begin{figure*}[t!]
\centering
\includegraphics[width=0.46\textwidth]{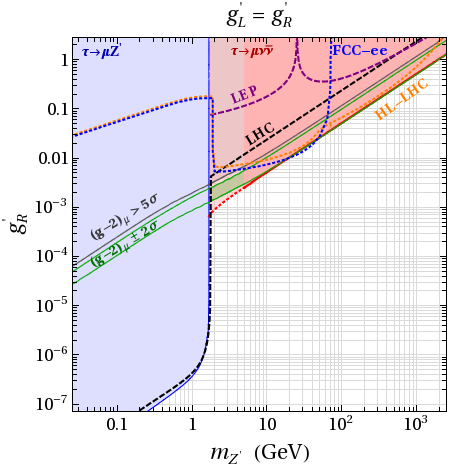} 
\includegraphics[width=0.46\textwidth]{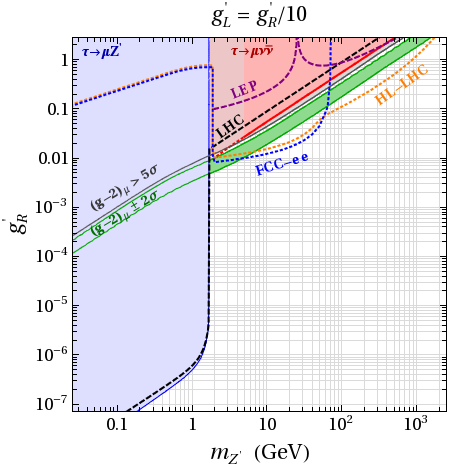}
\caption{Slices of $m_{Z^\prime}$ vs. $g^\prime_R$ parameter space. The left-handed coupling is set to $g_L^\prime = g_R^\prime$ in the left panel and $g_L^\prime = g_R^\prime / 10$ in the right panel. The green band is the $2\, \sigma$-preferred range by the $(g-2)_\mu$ anomaly, while the gray region is disfavored at $> 5 \, \sigma$ (see Section~\ref{sec:gm2}). The red region is excluded at $2\, \sigma$ by lepton flavor universality in tau decays (see Section~\ref{sec:lfv}). The blue region is excluded at 95\% CL by searches for the two-body decay $\tau \to \mu Z^\prime$ (see Section~\ref{sec:2body}). The black dashed curve shows the 95\% CL LHC exclusion from searches for leptonic $W$ decays (see Section~\ref{sec:lhc}) and the purple dashed curve shows the 95\% CL LEP exclusion from $Z$ coupling measurements (see Section~\ref{sec:lep}). The orange and blue dotted lines show the expected $3\sigma$ sensitivity in searches for the $\mu^\pm \mu^\pm \tau^\mp \tau^\mp$ final state at the high-luminosity LHC and 
at a future electron-positron collider running at the $Z$ pole (see Section~\ref{sec:4lepton}).}
\label{fig:3}
\end{figure*}

In Figs.~\ref{fig:2} and~\ref{fig:3} we show regions of parameter space that allow to address the $(g-2)_\mu$ discrepancy. The plot in Fig.~\ref{fig:2} shows the $g_L^\prime$ vs. $g_R^\prime$ plane for a fixed $Z^\prime$ mass $m_{Z^\prime} = 100$~GeV; the plots in Fig.~\ref{fig:3} show the $m_{Z^\prime}$ vs. $g_R^\prime$ plane for two choices of $g_L^\prime$, namely, $g_L^\prime = g_R^\prime$ (left) and $g_L^\prime = g_R^\prime /10$ (right). The green bands correspond to the $2\,\sigma$ preferred region from Eq.~(\ref{eq:gm2}). In the gray regions, the discrepancy is larger than $5\,\sigma$ which we consider to be excluded.
Note that both left-handed and right-handed couplings are required to explain the anomaly. 
Pure left-handed or pure right-handed couplings of the $Z^\prime$ necessarily enlarge the discrepancy in $(g-2)_\mu$, as seen from Eq.~\eqref{eq:g-2}, and hence, are not entertained here. Other constraints shown in Figs.~\ref{fig:2} and \ref{fig:3} are explained below. 

\section{Lepton Flavor Universality Violation in Tau Decays}\label{sec:lfv}

Constraints on our flavor violating $Z^\prime$ scenario can be derived from leptonic tau decays.
In the SM, the leptonic decays of the tau, $\tau^- \to \mu^- \nu_\tau \bar\nu_\mu$ and $\tau^- \to e^- \nu_\tau \bar\nu_e$, are mediated by the tree-level exchange of a $W$ boson.
Integrating out the $W$, we arrive at the following effective Hamiltonian describing the decays:
\begin{eqnarray}
 \mathcal H_\text{SM} & \ = \ & \frac{g_2^2}{2 m_W^2} (\bar \nu_\tau \gamma_\alpha P_L \tau)\sum_{\ell=e,\mu}(\bar \ell \gamma^\alpha P_L \nu_\ell) 
\,,
\end{eqnarray}
where
$g_2 = e / \sin\theta_W \simeq 0.65$ is the $SU(2)_L$ gauge coupling.
Due to lepton flavor universality of the weak interactions, the ratio of the branching ratios of the leptonic tau decays is close to unity. In the SM, the ratio can be predicted with extremely high accuracy~\cite{Pich:2013lsa}:
\begin{equation} \label{eq:tauSM}
 R_{\mu e}^\text{SM} = \frac{\text{BR}(\tau \to \mu \nu_\tau \bar\nu_\mu)_\text{SM}}{\text{BR}(\tau \to e \nu_\tau \bar\nu_e)_\text{SM}} = 0.972559 \pm 0.000005 \,,
\end{equation}
where the deviation from unity is almost entirely due to phase space effects.

\begin{figure}[t]
\centering
\includegraphics[width=5cm]{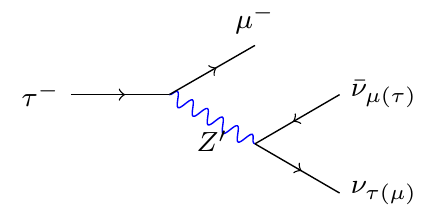}
\caption{Feynman diagram for the $Z'$ contribution to the lepton flavor universality violating tau decay.}
\label{fig:4}
\end{figure}

On the experimental side, the most precise measurement of this ratio comes from BaBar~\cite{Aubert:2009qj}. The PDG average~\cite{Agashe:2014kda} also includes less precise determinations from CLEO~\cite{Anastassov:1996tc} and ARGUS~\cite{Albrecht:1991rh}: 
\begin{equation} \label{eq:tauexp}
 R_{\mu e}^\text{PDG} \ = \ 0.979 \pm 0.004 \,.
\end{equation}
We observe a slight tension with the SM prediction at the level of $1.6\, \sigma$.
Combining Eqs.~(\ref{eq:tauSM}) and~(\ref{eq:tauexp}) we find
\begin{equation}\label{eq:taubound}
 \frac{R_{\mu e}}{R_{\mu e}^\text{SM}} - 1 \ = \ 0.0066 \pm 0.0041 \,.
\end{equation}
The tree level exchange of the considered flavor violating $Z^\prime$ cannot affect the $\tau \to e \nu_\tau \bar\nu_e$ decay. However, it does give additional contributions to the $\tau \to \mu \nu_\tau \bar\nu_\mu$ decay and induces the new tau decay mode $\tau \to \mu \nu_\mu \bar\nu_\tau$, as shown in Fig.~\ref{fig:4}. 
The decay $\tau \to \mu \nu_\mu \bar\nu_\tau$ is absent in the SM, but has exactly the same experimental signature as $\tau \to \mu \nu_\tau \bar\nu_\mu$. In the following we will therefore consider the sum of the two decay modes that we denote with $\tau \to \mu \nu \bar\nu$. As long as $m_{Z^\prime} \gg m_\tau$ the treatment of the $Z^\prime$ effect in terms of an effective Hamiltonian is valid and we find
\begin{eqnarray} 
 \mathcal H_{Z^\prime} & \ = \ &  -\frac{|g_L^\prime|^2}{m_{Z^\prime}^2} (\bar \mu \gamma^\alpha P_L \tau) (\bar \nu_\tau \gamma^\alpha P_L \nu_\mu) \nonumber \\
  && -\, \frac{g_R^\prime g_L^{\prime *}}{m_{Z^\prime}^2} (\bar \mu \gamma^\alpha P_R \tau) (\bar \nu_\tau \gamma_\alpha P_L \nu_\mu) \nonumber \\ 
  && -\, \frac{(g_L^{\prime})^2}{m_{Z^\prime}^2} (\bar \mu \gamma^\alpha P_L \tau) (\bar \nu_\mu \gamma_\alpha P_L \nu_\tau) \nonumber \\
  && -\, \frac{g_R^\prime g_L^\prime}{m_{Z^\prime}^2} (\bar \mu \gamma^\alpha P_R \tau) (\bar \nu_\mu \gamma_\alpha P_L \nu_\tau) \,.
\label{HamZp}
\end{eqnarray}
The Hamiltonian in Eq.~\eqref{HamZp} leads to the following correction to the lepton flavor universality ratio $R_{\mu e}$: 
\begin{equation} 
 \frac{R_{\mu e}}{R_{\mu e}^\text{SM}} \ = \ 1 + \frac{|g_L^\prime|^2}{g_2^2} \frac{4 m_W^2}{m_{Z^\prime}^2} + \left(\frac{|g_L^\prime g_R^\prime|^2}{g_2^4} + \frac{|g_L^\prime|^4}{g_2^4} \right) \frac{8 m_W^4}{m_{Z^\prime}^4} \,.
\label{eq:heavyZp}
\end{equation}
Note that our model can only increase the ratio $R_{\mu e}$ compared to the SM prediction.
Thus, the result in Eq.~(\ref{eq:taubound}) gives strong constraints on the $Z^\prime$ parameter space.
If we neglect the term that contains the right-handed $Z^\prime$ coupling, we find the following approximate constraint at the $2\,\sigma$ level
\begin{equation}
  \frac{m_{Z^\prime}}{|g_L^\prime|} \ \gtrsim \ 2~\text{TeV} ~.
\label{eq:mZpbound}
\end{equation}
Note that in the absence of $g_L^\prime$, the $Z^\prime$ does not couple to neutrinos and the constraint from lepton flavor universality in tau decays can be avoided.

The constraint~(\ref{eq:taubound}) is shown in Fig.~\ref{fig:2} in the $g_L^\prime$--$g^\prime_R$ plane in red, for a fixed $Z^\prime$ mass of $m_{Z^\prime} = 100$~GeV. Large values of $g_L^\prime$ are strongly constrained, leaving a compact region of $g_L^\prime$-$g_R^\prime$ parameter space, where the $(g-2)_\mu$ anomaly can be explained.
The dashed red lines show values of constant lepton flavor universality violation, i.e. $R_{\mu e}/R_{\mu e}^\text{SM} -1 = 1\%, 0.3\%, 0.1\%, 0.03\%$. 
Probing lepton flavor universality violation in tau decays down to a level of $0.1\%$ would allow us to conclusively test the entire remaining parameter space relevant for our explanation of the $(g-2)_\mu$ anomaly. This should be possible to achieve at Belle~2~\cite{Aushev:2010bq} with 50 ab$^{-1}$ luminosity, assuming that systematic uncertainties can be kept under control. 

For $Z^\prime$ masses of the order of the tau mass, the momentum transfer along the $Z^\prime$ propagator in Fig.~\ref{fig:4} has to be taken into account. In this case we find
\begin{eqnarray}
\frac{R_{\mu e}}{R_{\mu e}^\text{SM}} & \ = \ & 1 + \frac{|g_L^\prime|^2}{g_2^2} \frac{4 m_W^2}{m_{Z^\prime}^2}f\left(\frac{m_\tau^2}{m_{Z^\prime}^2}\right)+ \nonumber \\
&& + \left(\frac{|g_L^\prime g_R^\prime|^2}{g_2^4} + \frac{|g_L^\prime|^4}{g_2^4} \right) \frac{2 m_W^4}{m_{Z^\prime}^4} g\left(\frac{m_\tau^2}{m_{Z^\prime}^2}\right) ,
\label{eq:genZp}
\end{eqnarray}
with the functions
\begin{eqnarray}
 f(z) &=& \frac{2}{z^4} \left[ \frac{5}{6} z^3 + 2z^2 -2z - (1-z)^2(2+z)\log(1-z) \right] \,, \nonumber \\
 g(z) &=& \frac{2}{z^4} \left[ - z^3 -3z^2 +6z + 6(1-z)\log(1-z) \right] \,. \nonumber
\end{eqnarray}
In the limit $m_{Z^\prime} \gg m_\tau$, we have $\lim\limits_{z\to 0}f(z) =1$, $\lim\limits_{z\to 0}g(z)= 1$ and then Eq.~\eqref{eq:genZp} reduces to Eq.~(\ref{eq:heavyZp}).
Note that in the above expression we still neglected the muon mass.
Once the $Z^\prime$ mass comes close to the tau mass, such that $m_\tau - m_{Z^\prime} \sim m_\mu$ also the muon can no longer be treated massless. 
In our numerical analysis we take into account the muon mass.

In Fig.~\ref{fig:3} we show in red the regions in the $m_{Z^\prime}$--$g^\prime_R$ plane that are  excluded by the current experimental measurement of $R_{\mu e}$, for two choices of $g_L^\prime$. As expected, the constraint is strongest for large values of $g_L^\prime$. For a heavy $Z^\prime$ the constraint agrees well with the approximate bound in Eq.~(\ref{eq:mZpbound}). 
We observe that in the case $g_L^\prime = g_R^\prime$, the tau decays exclude entirely our explanation of the $(g-2)_\mu$ anomaly for any $Z^\prime$ mass larger than $m_\tau$.

\begin{figure}[t]
\centering
\includegraphics[width=0.46\textwidth]{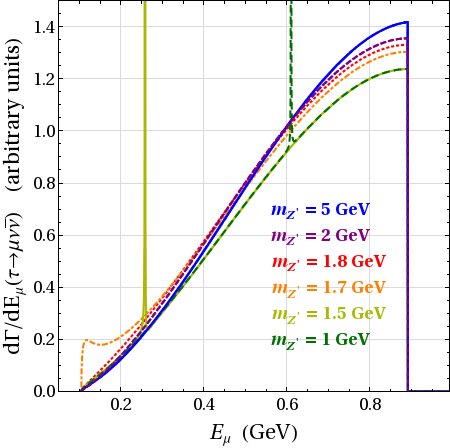}
\caption{Differential muon energy spectrum in the decay $\tau \to \mu \nu \bar\nu$ in the tau rest frame in the presence of a light $Z'$ with the indicated masses.}
\label{fig:5}
\end{figure}

A light $Z^\prime$ of the order of the tau mass not only affects the overall rate of the $\tau \to \mu \nu_\tau \bar\nu_\mu$ decay, but also modifies the muon energy spectrum. In Fig.~\ref{fig:5} we show the muon energy spectrum in the tau rest frame for various choices of the $Z^\prime$ mass. We set $g_L^\prime = g_R^\prime / 10$ and choose $g_R^\prime$ such that the $Z^\prime$ leads to a 10\% increase of the $\tau \to \mu \nu_\tau \bar\nu_\mu$ decay rate, i.e. $R_{\mu e} = 1.1 \times R_{\mu e}^\text{SM}$.
For a $Z'$ mass close to the tau mass, the $Z^\prime$ exchange leads to muons that tend to be slightly softer compared to that from the SM.
We caution the reader that a possible impact of the modified muon spectrum is not taken into account when deriving the bound in Fig.~\ref{fig:3}. A careful analysis of the experimental acceptances and efficiencies would be required to ascertain the robustness of the bound shown in the parameter region $ m_\tau - m_\mu < m_{Z^\prime} \lesssim$ few~GeV (shown by the dotted red curve), which is beyond the scope of this work. 

A more detailed study of the $Z'$ effect might include a Michel parameter analysis for tau decays~\cite{Pich:2013lsa, Tobe:2016qhz}. Moreover, a study of the tau polarization via its decays may be useful to probe differences from the SM, induced by the $Z'$ effect.

\section{Two-body Tau Decays} \label{sec:2body}

If the $Z^\prime$ mass is smaller than the difference of tau and muon mass, $m_{Z^\prime} < m_\tau - m_\mu$ the two body decay $\tau \to \mu Z^\prime$ opens up kinematically. This is illustrated for the two cases $m_{Z^\prime} = 1.5$~GeV and $m_{Z^\prime} = 1$~GeV by the peaks in the muon energy spectrum shown in Fig.~\ref{fig:5}. In this region of parameter space the only available decay mode of the $Z^\prime$ is into neutrinos. Direct searches for the decay $\tau \to \mu +$missing energy can then be used to constrain the $Z^\prime$ parameter space.

The $\tau \to \mu Z^\prime$ decay width reads
\begin{align}
\Gamma_{\mu Z'} &  =  \frac{m_\tau^3}{32\pi m_{Z^\prime}^2} \Bigg[ \left( |g_L^\prime|^2 + |g_R^\prime|^2 \right) \bigg\{ \left( 1 + \frac{2m_{Z^\prime}^2}{m_\tau^2} \right)\left( 1 - \frac{m_{Z^\prime}^2}{m_\tau^2} \right) \nonumber \\
&  - \frac{m_\mu^2}{m_\tau^2} \left( 2 - \frac{m_{Z^\prime}^2}{m_\tau^2} -\frac{m_\mu^2}{m_\tau^2}\right) \bigg\} - 12\, \text{Re}(g_L^\prime g_R^{\prime *}) \frac{m_\mu}{m_\tau} \frac{m_{Z^\prime}^2}{m_\tau^2} \Bigg] \nonumber \\
& \times \sqrt{\left( 1 - \frac{m_{Z^\prime}^2}{m_\tau^2} \right)^2 - \frac{m_\mu^2}{m_\tau^2} \left( 2 + \frac{2m_{Z^\prime}^2}{m_\tau^2} -\frac{m_\mu^2}{m_\tau^2}\right)  }\,.
\label{eq:muzp}
\end{align}
Neglecting terms that are suppressed by the muon mass, this can be simplified to 
\begin{eqnarray}
 \Gamma_{\mu Z^\prime} & \ \simeq \ & \frac{m_\tau^3}{32\pi m_{Z^\prime}^2} \left( |g_L^\prime|^2 + |g_R^\prime|^2 \right) \nonumber \\ 
 && \times \left( 1 + \frac{2m_{Z^\prime}^2}{m_\tau^2} \right)\left( 1 - \frac{m_{Z^\prime}^2}{m_\tau^2} \right)^2 \,.
\end{eqnarray}
In our numerical analysis we keep muon mass effects and use Eq.~\eqref{eq:muzp}. 

Searches for the two body decay $\tau \to \mu \phi$ by ARGUS~\cite{Albrecht:1995ht}, where $\phi$ is an unobservable particle, directly apply to our case; they give bounds on the corresponding branching ratio for masses up to $1.6$~GeV.
The region of $Z^\prime$ parameter space that is excluded by this search is shown in Fig.~\ref{fig:3} in blue. 
The bound from $\tau \to \mu Z^\prime$ is remarkably strong, and as a result, our explanation of the  $(g-2)_\mu$ is entirely excluded for $m_{Z^\prime} \lesssim m_\tau - m_\mu$ by orders of magnitude, independent of the relative size of $g_L^\prime$ and $g_R^\prime$.

\section{LHC Constraints} \label{sec:lhc}

The direct LHC constraints on $Z'$ from simple resonance searches like $pp\to Z'\to \ell^+\ell^-$ and $pp\to Z'\to jj$ are not applicable in our case, since the $Z'$ does not couple to quarks at the tree level. Moreover, the flavor-violating $Z'$ searches at the LHC have only focused on the $e\mu$ channel so far~\cite{atlas:emu, CMS:2016dfe}.
Nevertheless, we can derive LHC constraints on the $\mu\tau$ coupling from the leptonic decays of the $W$ boson, since $pp\to W\to \mu \nu$ will also get a contribution from $pp\to W\to \tau \nu$, followed by the $Z'$-mediated decay of tau, as shown in Fig.~\ref{fig:4}. This will lead to an isolated muon and three neutrinos in the final state, where the neutrinos will be registered in the LHC detectors simply as missing energy, without any information on their number or flavor content. So we can use the constraints derived from this channel in our case, as long as the missing energy criterion $E_T^{\rm miss}>25$ GeV used in the corresponding $\mu\nu$ search at $\sqrt s=13$ TeV LHC~\cite{Aad:2016naf} is satisfied.

To check this, we implemented our model Lagrangian~\eqref{lagZp} into MadGraph5~\cite{Alwall:2014hca} for event generation with CT14NNLO PDFs~\cite{Dulat:2015mca}, used PYTHIA 6.4~\cite{Sjostrand:2006za} for showering and hadronization, and DELPHES 3~\cite{deFavereau:2013fsa} for a fast detector simulation. We find that most of our $W \to \mu 3\nu$ signal events pass the event selection cuts of Ref.~\cite{Aad:2016naf} for a wide range of $Z'$ masses of interest. Here we have used the narrow-width approximation and have written down the $Z'$-induced cross section as 
\begin{eqnarray}
&& \sigma(pp\to W\to \tau\nu_\tau \to \mu\nu_\tau\nu_\mu\nu_\tau) \ =  \ \sigma_{\rm SM}(pp\to W\to \tau\nu_\tau)\nonumber \\
&& \qquad \qquad \qquad \qquad \qquad \qquad \quad  \times\:  {\rm BR}(\tau \to \mu\nu_\tau\nu_\mu\nu_\tau)\, .
\end{eqnarray}
For $m_{Z'}> m_\tau-m_\mu$, we use the following expression for the width of the 3-body decay $\tau^\pm \to \mu^\pm Z'^*\to \mu^\pm \nu\bar{\nu}$: 
\begin{align}
\Gamma_{\mu\nu\bar{\nu}} \ = \ |g_L'|^2(|g_L'|^2+|g_R'|^2)\frac{m_\tau^5}{768 \pi^3 m_{Z'}^4} a\left(\frac{m_\mu^2}{m_\tau^2}\right) \, ,
\label{3body}
\end{align}
where $a(x)=1-8x+8x^3-x^4-12x^2\log x$ and we included both channels $\tau \to \mu \nu_\mu \bar{\nu}_\tau$ and $\tau \to \mu \nu_\tau \bar{\nu}_\mu$. For $m_{Z'}< m_\tau-m_\mu$, we use the 2-body decay formula as in Eq.~\eqref{eq:muzp}. 
Comparing the measured value of the $pp\to W\to \mu \nu$ cross section $\sigma_{\rm exp} =20.64\pm 0.70$~nb~\cite{Aad:2016naf} with the SM NNLO prediction~\cite{Catani:2009sm} $\sigma_{\rm SM} =20.08\pm 0.66$~nb obtained using CT14NNLO PDFs~\cite{Dulat:2015mca}, we derive 95\% CL upper limits on the $Z'$ couplings, as shown in Figs.~\ref{fig:2} and \ref{fig:3} by the black dashed curves.  We find that the LHC constraints are weaker than the low-energy constraints directly derived from $\tau$ decay.  Future run-II LHC data, as well as the high-luminosity phase~\cite{Rossi:2011zc} and/or a future 100 TeV collider~\cite{Golling:2016gvc} will perhaps be able to probe a large portion of the allowed parameter space in Figure~\ref{fig:3}, if the systematics and the SM theory uncertainty could be improved.

\section{LEP Constraints} \label{sec:lep}

\begin{figure}[t]
\centering
\includegraphics[width=5cm]{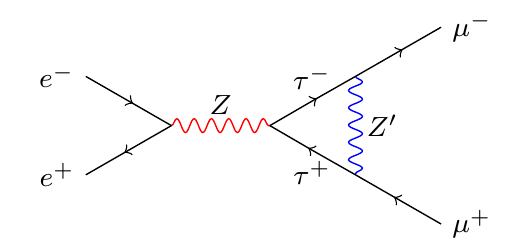}
\caption{Example Feynman diagram for the one-loop correction to the $Z$-decay vertex due to LFV $Z'$ interactions.}
\label{fig:lep}
\end{figure}

While our $Z'$ does not couple to electrons and quarks at tree level, it can contribute, however, to the processes $e^+e^-\to Z\to \mu^+\mu^-, \tau^+\tau^-$, and $\nu \bar\nu$ via one-loop diagrams involving $Z'$, as shown in Fig.~\ref{fig:lep}.
Measurements of the SM $Z$ couplings~\cite{ALEPH:2005ab} to muons, taus and neutrinos can therefore be used to set constraints on the $Z'$ parameter space. 
We find the following modifications of the $Z$ couplings due to the $Z'$ loop:
\begin{eqnarray}
\frac{g_{L\tau}}{g_{Le}} \ \simeq \ \frac{g_{L\mu}}{g_{Le}} & \ \simeq \ & 1 + \frac{|g_L^\prime|^2}{16\pi^2} ~\mathcal{K}(m_Z^2/m_{Z^\prime}^2) \, , \\ 
\frac{g_{R\tau}}{g_{Re}} \ \simeq \ \frac{g_{R\mu}}{g_{Re}} & \ \simeq \ & 1 + \frac{|g_R^\prime|^2}{16\pi^2} ~\mathcal{K}(m_Z^2/m_{Z^\prime}^2) \, , \\ 
\frac{g_{L\nu}}{g_{Re}- g_{Le}} & \ \simeq \ & 1 + \frac{2}{3}\frac{|g_L^\prime|^2}{16\pi^2} ~\mathcal{K}(m_Z^2/m_{Z^\prime}^2) \, ,
\end{eqnarray}
with the loop function $\mathcal{K}$~\cite{Haisch:2011up}
\begin{eqnarray}
 \mathcal{K}(x) &=& - \frac{4+7x}{2x} + \frac{2+3x}{x}\log x  \nonumber \\
 &-& \frac{2(1+x)^2}{x^2}\big[\log x \log(1+x) + \text{Li}_2(-x) \big] \, ,
\label{loop}
\end{eqnarray}
where $\text{Li}_2(x) = - \int_{0}^x dt \log(1-t)/t$ is the di-logarithm.
In the above expressions, we use the electron couplings $g_{Le}$ and $g_{Re}$ as convenient normalization, as they are not affected by $Z'$ loops. 

The combined experimental results for the $Z$ couplings from LEP and SLD read~\cite{ALEPH:2005ab}
\begin{eqnarray}
 g_{L\nu} & \ = \ & +0.5003 \pm 0.0012 \,,  \\
 g_{Le} & \ = \ & -0.26963 \pm 0.00030 \,,  \\
 g_{L\mu} & \ = \ & -0.2689 \pm 0.0011 \,,  \\
 g_{L\tau} & \ = \ & -0.26930 \pm 0.00058 \,,  \\
 g_{Re} & \ = \ & +0.23148 \pm 0.00029 \,,  \\
 g_{R\mu} & \ = \ & +0.2323 \pm 0.0013 \,,  \\
 g_{R\tau} & \ = \ & +0.23274 \pm 0.00062 \,,
\end{eqnarray}
with the error correlation matrix
\begin{equation}
 \rho =   
{\small \begin{pmatrix} 
 1 & -0.52 & 0.12 & 0.22 & 0.37 & -0.06 & -0.17 \\
 -0.52 & 1 & -0.11 & -0.07 & 0.29 & -0.06 & 0.04 \\
 0.12 & -0.11 & 1 & 0.07 & -0.07 & 0.90 & -0.04 \\
 0.22 & -0.07 & 0.07 & 1 & 0.01 & -0.03 & 0.44 \\
 0.37 & 0.29 & -0.07 & 0.01 & 1 & -0.09 & -0.03 \\
 -0.06 & -0.06 & 0.90 & -0.03 & -0.09 & 1 & 0.04 \\
 -0.17 & 0.04 & -0.04 & 0.44 & -0.03 & 0.04 & 1 
 \end{pmatrix}.}
\end{equation}
To derive bounds on the $Z'$ couplings and mass we perform a simple $\chi^2$ fit, setting the electron couplings $g_{Le}$ and $g_{Re}$ to the measured values. The resulting constraint is shown in the plots of Figs.~\ref{fig:2} and~\ref{fig:3} as dashed purple curves. Above the curves $\Delta \chi^2 > 4$, corresponding to a $95\%$ CL exclusion. The constraint vanishes around $m_{Z'}=25$~GeV, where the loop function~\eqref{loop} has a zero crossing.
We observe that the LEP constraint is generically weaker than the constraint obtained from the tree-level leptonic tau decays. Similarly, the constraints obtained from the modifications to the $W$ and $Z$ total widths due to the $Z'$ effects~\cite{Laha:2013xua} are weaker than those in the whole parameter space of interest, and therefore, are not shown in Fig.~\ref{fig:3}.

\section{Future Collider Signatures}\label{sec:4lepton}

\begin{figure}[t]
\centering
\includegraphics[width=5cm]{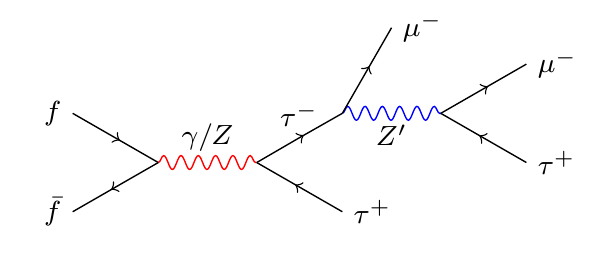}
\caption{A striking collider signature of our LFV $Z'$ scenario. This is applicable to both lepton and hadron colliders (depending on whether the initial state fermion $f$ is a SM charged lepton or quark). There exists a similar diagram with an intermediate muon, which is not shown here, but included in our calculation.}
\label{fig:col}
\end{figure}

The leptophilic $Z'$ scenarios have characteristic multi-lepton signatures at both lepton and hadron colliders~\cite{delAguila:2014soa, Elahi:2015vzh}.  A particularly interesting signal in our LFV $Z'$ scenario is the 4-lepton final state with two same-sign muons and taus at the LHC, i.e. 
\begin{align}
pp \ \to \ \mu^\pm \tau^\mp Z'^{(*)} \ \to \ \mu^\pm \mu^\pm \tau^\mp \tau^\mp \, ,
\label{lhc}
\end{align}
as shown in Fig.~\ref{fig:col}. This signal is very clean and effectively background-free. Through this process, one might also be able to determine the $Z^\prime$ mass for $m_{Z'}>m_\tau+m_\mu$, when the $Z'$ in Fig.~\ref{fig:col} goes on-shell and one of the $\mu\tau$ pairs will have an invariant mass at $m_{Z'}$. Although the tau reconstruction poses some practical challenges, Eq.~\eqref{lhc} could provide a `smoking gun' signal for our $Z'$ scenario at the LHC.

We simulate the process \eqref{lhc} to estimate the sensitivity reach at the $\sqrt s=14$ TeV LHC. The parton level events are generated using MadGraph5~\cite{Alwall:2014hca}, which are then fed to PYTHIA 6.4~\cite{Sjostrand:2006za} for showering and hadronization, and DELPHES 3~\cite{deFavereau:2013fsa} for a fast detector simulation. We impose the basic trigger cuts following a previous analysis for light $Z'$ searches in the $pp\to Z\to 4\mu$ channel~\cite{Elahi:2015vzh}: 
\begin{itemize}
 \item[(i)] the leading lepton must satisfy the transverse momentum cut $p_T>20$ GeV, while the sub-leading leptons are required to satisfy a milder cut $p_T>15$ GeV;
 \item[(ii)] all the four leptons must satisfy the pseudo-rapidity $|\eta|<2.7$ and the isolation cut $\Delta R>0.1$.
\end{itemize}
These values are set to be as inclusive as possible for an optimistic analysis.

Since we are interested in the final states with same-sign muon pairs, we select the hadronic decay mode of the taus. In the SM, each tau decays hadronically with a probability of $\sim$ 
65\%, producing a tau-jet mostly containing neutral and charged pions. In our case with a pair of taus in the final state, 42\% of the events will contain two tau-jets. The hadronic tau decays have low charged track multiplicity (one or three prongs) and a relevant fraction of the electromagnetic energy deposition due to photons coming from the decay of neutral pions. Moreover, when the momentum of the tau is large compared to its mass, the tau-jets will be highly collimated and produce localized energy deposit in the electromagnetic and hadronic calorimeters. These characteristics can be exploited to enhance the identification of hadronic tau decays~\cite{Bagliesi:2007qx}. We have assumed an optimistic value of 70\% for the tau-tagging efficiency in our analysis. Since the SM background is negligible for the same-sign di-lepton pairs $\mu^\pm \mu^\pm \tau^\mp \tau^\mp$, we can simply estimate the signal sensitivity as ${\cal N}=S/\sqrt{S+B}\simeq \sqrt{{\cal L}\sigma_{\rm signal}}$, where ${\cal L}$ is the 
integrated luminosity and $\sigma_{\rm signal}$ is the signal cross section times efficiency, as obtained from our detector simulation. 

Our results for the $3\, \sigma$ sensitivity reach (corresponding to ${\cal N}>3$) in the high-luminosity phase of the LHC with 3 ab$^{-1}$ integrated luminosity are shown in Figs.~\ref{fig:2} and \ref{fig:3} by the orange curves. The LHC sensitivity gets weaker for a very light $Z^\prime$ with mass $m_{Z'}<m_\tau+m_\mu$, since both the intermediate tau as well as $Z'$ in Fig.~\ref{fig:col} are off-shell in this case. 
The small bump around $Z$-mass is because the $Z$ also becomes off-shell for $m_{Z'}>m_Z-m_\mu$.
Overall, we find that for $m_{Z'}>m_\tau+m_\mu$, the HL-LHC has good sensitivity to large parts of the $(g-2)_\mu$-favored region. The LHC sensitivity again becomes weaker for a very heavy $m_{Z'}>2$ TeV or so, simply due to the kinematic suppression. A future $\sqrt s=100$ TeV $pp$ collider could extend our LFV $Z'$ sensitivity reach to the multi-TeV range.

The collider sensitivity can be further improved for $m_{Z'}<m_Z$ by considering a lepton collider operating at the $Z$-pole, i.e. with $\sqrt s=m_Z$. As an example, we consider a next generation $e^+e^-$ $Z$ factory such as the FCC-ee, and simulate the process (cf. Fig.~\ref{fig:col})
\begin{align}
e^+e^- \ \to \ Z \ \to \ \mu^\pm \tau^\mp Z'^{(*)} \ \to \ \mu^\pm \mu^\pm \tau^\mp \tau^\mp 
\label{ilc}
\end{align}
for $m_{Z'}<m_Z$ using the procedure outlined above. Our results are shown in Fig.~\ref{fig:3} by the blue curves for the maximum achievable integrated luminosity of 2.6 ab$^{-1}$ at FCC-ee~\cite{d'Enterria:2016yqx}. We find that the sensitivity can be improved by a factor of 2-3, thus covering almost the entire $(g-2)_\mu$-favored region for $m_{Z'}<m_Z$.

Our $Z'$ scenario can in principle also affect the SM Higgs decays. First of all, the $h\to \mu^+\mu^-$ decay will receive a one-loop correction due to the LFV $Z'$ interactions. Although it is enhanced by the tau Yukawa coupling, 
due to the loop suppression factor, and given that BR$(h\to \mu\mu)=2.19\times 10^{-4}$ in the SM~\cite{Heinemeyer:2013tqa}, the deviation due to the $Z'$-loop correction is extremely difficult to be observed at the HL-LHC or even at a dedicated Higgs factory.

The $Z'$ interactions could also induce a LFV decay of $h\to \tau^+\tau^-\to \mu^\pm \tau^\mp Z'$, where the $Z'$ goes undetected for a sufficiently small $m_{Z'}$. However, for the allowed range of masses and couplings in Fig.~\ref{fig:3}, this effect is again small and easily compatible with the LHC searches for $h \to \mu\tau$ that imply BR$(h\to \mu\tau) \lesssim 1.5\%$~\cite{CMS:2016qvi,Aad:2016blu}.  

\section{Signal at Neutrino Telescopes} \label{sec:icecube}

\begin{figure}[t]
\centering
\includegraphics[width=4cm]{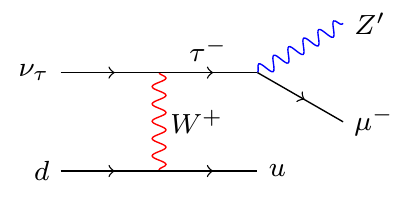}
\includegraphics[width=4cm]{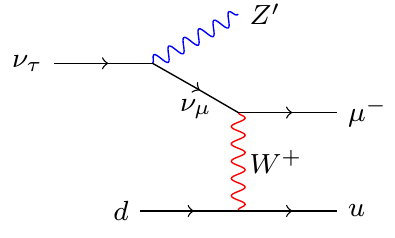}
\caption{$Z'$ contributions to the charged-current neutrino-nucleon interactions. Similar diagrams for incident neutrinos of muon flavor, as well as for antineutrinos and also for neutral-current interactions, are not shown here.}
\label{fig:6}
\end{figure}

In this section, we briefly discuss a complementary way to test our LFV $Z'$ hypothesis using ultra-high energy (UHE) neutrinos in large volume neutrino telescopes like IceCube and KM3NeT. First, we note that the $Z'$ interactions in our model induce new channels for the neutrino-nucleon interactions, as illustrated in Fig.~\ref{fig:6} for the charged-current (CC) process. For $m_{Z'}> m_\tau+m_\mu$, they could potentially give rise to a novel signature with simultaneous muon and tau events coming from the $Z'\to \mu\tau$ decay. However, it turns out that in presence of both left and right-handed $Z'$ couplings to charged leptons, as required for the $(g-2)_\mu$ explanation, there is a destructive interference between the two diagrams in Fig.~\ref{fig:6}, which leads to a cross section too small to be ever observed. Moreover, the stringent limits on the $Z'$ couplings from tau decays (cf. Figs.~\ref{fig:2} and \ref{fig:3}) necessarily imply that even if we disregard the $(g-2)_\mu$ favored 
region by taking $g_R'\gg g_L'$, the total cross section for the processes in Fig.~\ref{fig:6} is still small, as compared to that of the SM CC interaction. To give an example, for a benchmark point with $m_{Z'}=2$ GeV, $g_R'=0.02$ and $g_L'=0.0004$ which satisfies the $\tau\to \mu\nu\bar{\nu}$ constraint, we find the total cross section for the processes shown in Fig.~\ref{fig:6} (including the antineutrino initial states) for an incoming neutrino energy $E_\nu= 1$ PeV to be $1.54\times 10^{-38}~{\rm cm}^2$, as compared to the corresponding SM CC cross section of $7.3\times 10^{-34}~{\rm cm}^2$, both calculated using the CT14NNLO PDFs~\cite{Dulat:2015mca}. It is difficult to measure such a small cross section at IceCube even with large statistics, since it will be overshadowed by various uncertainties in the incoming neutrino flux, flavor composition, and parton distribution functions (see e.g.~\cite{Chen:2013dza, Chen:2014gxa, Vincent:2016nut}).  

A better possibility to detect a light $Z'$ at IceCube might be through its effect on neutrino-neutrino scattering due to on-shell $Z'$ production. In fact, the resonant absorption of UHE neutrinos by the cosmic neutrino background (C$\nu$B)~\cite{Weiler:1982qy, Weiler:1983xx, Weiler:1997sh, Roulet:1992pz, Yoshida:1996ie, Fargion:1997ft} in the presence of a light mediator has been invoked~\cite{Ibe:2014pja, Araki:2014ona, Araki:2015mya, Kamada:2015era, DiFranzo:2015qea} to explain the apparent energy gap in the IceCube neutrino data~\cite{Aartsen:2013jdh, Aartsen:2014gkd, Aartsen:2015zva} just below the PeV deposited energy bin. However, this scenario works only for an MeV-scale $Z'$, which is unfortunately ruled out in our model due to the $\tau \to \mu Z'$ constraint. For a higher $Z'$ mass, the incoming neutrino energy required to observe a resonance feature at the IceCube will be shifted upwards: 
\begin{align}
E_{\nu}^{\rm res} \ = \ \frac{m_{Z'}^2}{2m_\nu(1+z)} \, ,
\label{res}
\end{align}
where $z$ is the redshift parameter at which the scattering occurs (typically taken to be the source redshift),\footnote{The redshift factor $(1+z)$ in Eq.~\eqref{res} is due to the fact that the energy $E_{\nu_i}^s$ of the cosmic neutrino $\nu_i$ at the source position $z$ is $(1+z)$ times the energy $E_{\nu}$ measured at IceCube in an expanding Universe.} $m_\nu$ is the mass of the target C$\nu$B, which is assumed to be larger than the effective temperature of the thermal distribution of the C$\nu$B, $T_{\nu}=1.7\times 10^{-4}(1+z)$ eV.\footnote{If the lightest neutrino is nearly massless, $m_\nu$ in Eq.~\eqref{res} should be replaced with the thermally averaged momentum $\langle p_\nu \rangle = \frac{7\pi^4 T_\nu}{180\zeta(3)}\approx 3.15 T_\nu$.} 

The total cross section for $\nu_i\bar{\nu}_j\to Z'\to f\bar{f'}$, where $\{i,j\}=\{\mu,\tau\}$ and $\{f,f'\} = \{\nu_\mu,\nu_\tau\}$ or $\{\mu,\tau\}$ (with $i\neq j \,, f\neq f'$), is given by
\begin{align}
\sigma(s) \ = \ \frac{1}{6\pi}|g_L'|^2(2|g_L'|^2+|g_R'|^2)\frac{s}{(s-m_{Z'}^2)^2+m_{Z'}^2\Gamma_{Z'}^2} \, ,
\label{cross}
\end{align}
where $s$ is the squared center of mass energy and $\Gamma_{Z'}$ is the total width of the $Z'$. Here we have ignored the $t$-channel contribution for the $\nu\bar{\nu}$ final state, as it is highly suppressed relative to the $s$-channel resonance. Also we have assumed $s \gg (m_\tau+m_\mu)^2$. For $m_{Z'}>m_\tau+m_\mu$, there are two decay modes of $Z'\to \nu_{\mu(\tau)}\bar{\nu}_{\tau(\mu)},\: \mu^\pm \tau^\mp$, with the corresponding decay widths given by
\begin{align}
\Gamma_{\nu_\mu\bar{\nu}_\tau} & =  \frac{|g_L'|^2 m_{Z'}}{24\pi}\, ,\\
\Gamma_{\mu^-\tau^+}  & =  \frac{m_{Z'}}{24\pi} \beta \tilde\beta \left[ \tilde \beta^2 \left(3 -\beta^2 \right) C_V^2 + \beta^2 \left(3 -\tilde\beta^2 \right) C_A^2 \right] \, ,
\end{align}
where $\beta=\sqrt{1-\frac{(m_\tau+m_\mu)^2}{m_{Z'}^2}}$, $\tilde\beta=\sqrt{1-\frac{(m_\tau-m_\mu)^2}{m_{Z'}^2}}$ and $C_{V,A}$ are defined below Eq.~\eqref{gm2}. The total decay width of $Z'$ is then given by $\Gamma_{Z'}=2\: (\Gamma_{\nu_\mu\bar{\nu}_\tau}+\Gamma_{\mu^-\tau^+})$, taking into account two possibilities for each decay mode.  
 \begin{figure}[t]
\centering
\includegraphics[width=8cm]{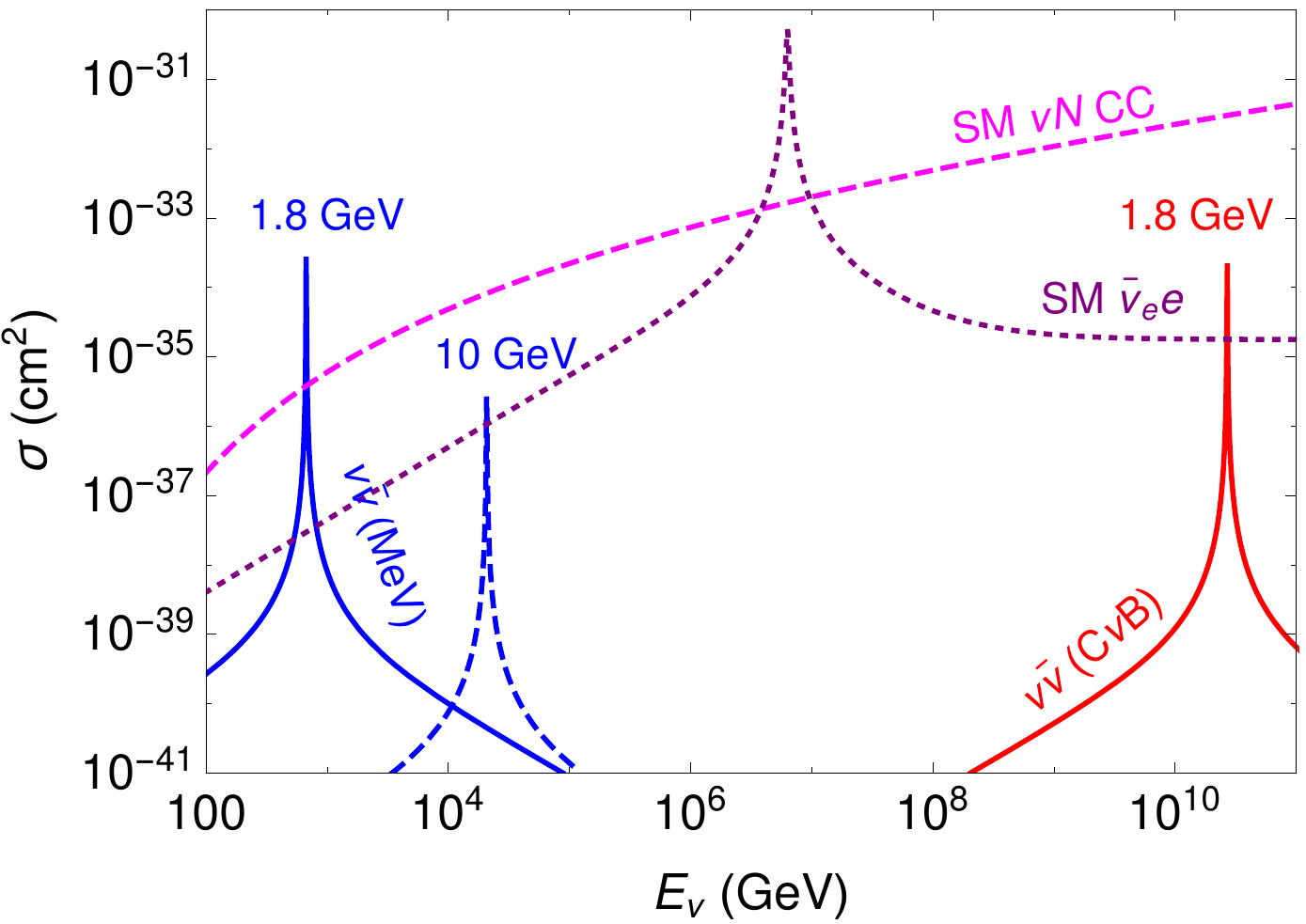}
\caption{Cross section for $\nu_i\bar{\nu}_j\to Z'\to f\bar{f'}$ as a function of the energy of one of the initial state neutrinos. For the second neutrino $\nu_j$, we consider two cases: C$\nu$B (red solid curve) and supernova neutrinos with MeV energy (blue solid and dashed curves). The numbers above the peaks show the $Z'$ mass. For comparison, we also show the SM neutrino-nucleon CC and $\bar{\nu}_e e$ cross sections.}
\label{fig:7}
\end{figure}

The cross section~\eqref{cross} is plotted in Fig.~\ref{fig:7} as a function of the energy of the incoming UHE neutrino $\nu_i$ for three different cases, depending on the energy of the other neutrino $\nu_j$. First, we consider the C$\nu$B for which the effective temperature $T_\nu$ is smaller than at least two of the light neutrino masses, so $s=2m_\nu E_\nu$. From Eq.~\eqref{res}, it is clear that for $m_{Z'}$ above the tau mass, the resonance will occur at very high energies well beyond the energy scale currently being probed at the IceCube. 
For an illustration, we choose a benchmark point from Fig.~\ref{fig:3} (right panel) satisfying all the constraints: $m_{Z'}=1.8$ GeV, $g_R'=0.01$, and $g_L'=g_R'/10$ and take the light neutrino mass $m_\nu=\sqrt{\Delta m_{\rm atm}}\simeq 0.05$ eV and a typical source redshift $z=0.2$. For this benchmark, we find the resonance energy to be at $2.7\times 10^{10}$ GeV,\footnote{For comparison, the SM $Z$ resonance occurs at $6.9\times 10^{13}$ GeV for $z=0.2$.} as shown by the red solid curve in Fig.~\ref{fig:7}. The other two light neutrino mass eigenstates will induce similar peaks at different energies, depending on their mass hierarchy.
For comparison, we also show the SM neutrino-nucleon CC and $\bar{\nu}_e e$ cross sections, with the latter having the Glashow resonance~\cite{Glashow:1960zz} at 6.3 PeV. In spite of the resonance enhancement, the $\nu\bar{\nu}$ cross section turns out to be much smaller than the SM $\nu N$ cross section. 

In order to check the condition under which the UHE neutrinos $\nu_i$ will likely have at least one interaction with the C$\nu$B during their entire journey from the source to Earth, we calculate their mean free path (MFP), given by  
\begin{align}
\lambda(E_\nu, z) & \ = \left[\int\frac{d^3p}{(2\pi)^3}\frac{1}{\exp{\{{\mathbf{p}/T_\nu(1+z)}\}}+1}\sigma(E_{\nu_i}^s,\mathbf{p}) \right]^{-1} \nonumber \\
& \ = \  \frac{1}{n_\nu \sigma(E_\nu)} \, ,  
\end{align} 
where $n_\nu = \frac{3}{4\pi^2}\zeta(3)T_\nu^3\simeq 56(1+z)^3~{\rm cm}^{-3}$ is the number density of the C$\nu$B (for each flavor) and $\sigma$ is given by Eq.~\eqref{cross}. The MFP will be the minimum at the resonance energy which corresponds to the maximum cross section. The survival rate of the high-energy neutrino $\nu_i$ travelling from the source at $z$ to Earth (at $z=0$) is then given by 
\begin{align}
P(E_\nu,z) \ = \ \exp{\left[-\int_0^z dz'\frac{1}{\lambda(E_\nu,z')}\frac{dL}{dz'}\right]} \, ,
\end{align}
where $dL/dz=c/(H_0\sqrt{\Omega_m(1+z)^3+\Omega_\Lambda})$, $c$ is the speed of light in vacuum, and the present best-fit values of the cosmological parameters in a $\Lambda$CDM Universe are $H_0=100h~{\rm km \, s}^{-1}{\rm Mpc}^{-1}$ with $h=0.678$, the matter energy density $\Omega_m=0.308$ and the dark energy density $\Omega_\Lambda=0.692$~\cite{Ade:2015xua}. Thus, if the traveling distance of the UHE neutrinos is larger than the MFP, they will be attenuated by the $C\nu$B and their survival rate will have a `dip' at the resonance energy. This will lead to a characteristic absorption feature in the UHE neutrino energy spectrum. For the benchmark discussed above, we find $\lambda(E_\nu^{\rm res})\simeq 6$ kpc, which means that all extragalactic sources like gamma-ray bursts and active galactic nuclei (with typical distances of Mpc or larger) or even far-away galactic sources like supernova remnants could in principle show an absorption feature in their neutrino spectrum due to the presence of a light $Z'$.

The resonance energy could be lowered significantly if we consider interactions of the high-energy neutrinos with other relativistic neutrinos naturally available, e.g. supernovae neutrinos (after they have oscillated into muon and tau flavors) which have a typical energy $E'$ in the MeV range~\cite{Formaggio:2013kya}. In this case, the center-of-mass energy of the system is $s=4E_\nu E'$, and the resonance condition~\eqref{res} gets modified to $E_\nu^{\rm res}=m_{Z'}^2/4E'(1+z)$, independent of the light neutrino mass, thus lowering the resonance energy down to the TeV scale. This is illustrated in Fig.~\ref{fig:7} for two choices of $m_{Z'}=1.8$ GeV (blue solid curve) and 10 GeV (blue dashed curve). Below the TeV scale, it will be difficult to observe the resonance feature, since it will be swamped by the atmospheric neutrino background. The neutrino number density at the supernova core surface is much larger, e.g. $\gtrsim 10^{34}~{\rm cm}^{-3}$ for SN1987A~\cite{Hirata:1987hu}. Hence, the MFP can be 
much smaller, thus allowing for the possibility of observing the absorption feature from both galactic and extragalactic sources, provided the incoming high-energy neutrinos encounter a supernova core collapse en route to Earth. The likelihood of such an arrangement somewhat depends on the origin of the high-energy neutrino source, and cannot be excluded at the moment.

The LFV interactions could also alter the ratio of astrophysical neutrino flavors at detection on Earth from the standard expectation of $(\nu_e:\nu_\mu:\nu_\tau)=(1:1:1)$. The detailed predictions for the event rate and the track-to-shower ratio will depend on many parameters, including the source neutrino flux normalization and spectral index, redshift, as well as the PDF uncertainties, but in spite of all these  uncertainties, the anomalous features could plausibly be measured~\cite{Barenboim:2004di, D'Olivo:2005uh, Lunardini:2013iwa, Ioka:2014kca, Ng:2014pca, Blum:2014ewa} by IceCube or next generation neutrino telescopes like IceCube-Gen2, thereby opening a new era of `cosmic neutrino spectroscopy'.

\section{Conclusion} \label{sec:concl}

We have discussed a simple new physics interpretation of the long-standing anomaly in the muon anomalous magnetic moment in terms of a purely flavor off-diagonal $Z'$ coupling only to the muon and tau sector of the SM. We have discussed the relevant constraints from lepton flavor universality violating tau decays for $m_{Z'}>m_\tau$ and from $\tau\to \mu ~+$ invisibles decay for $m_{Z'}<m_\tau$, as well as the latest LHC constraints from $W\to \mu\nu$ searches. 
We find that for a $Z'$ lighter than the tau, the low-energy tau decay constraints rule out the entire $(g-2)_\mu$ allowed region by many orders of magnitude. 
However, a heavier $Z'$ solution to the $(g-2)_\mu$ puzzle is still allowed, provided the $Z'$ coupling to the charged leptons has both left- and right-handed components, and the right-handed component is larger than the left-handed one. The deviations from lepton flavor universality in the tau decays predicted in this model can be probed at Belle 2, while a large part of the $(g-2)_\mu$ allowed region can be accessed at future colliders such as the high-luminosity LHC and/or an $e^+e^-$ $Z$-factory such as FCC-ee. The on-shell production of $Z'$ in high-energy neutrino interactions with either cosmic neutrino background or with other natural neutrino sources such as supernova neutrinos could lead to characteristic absorption features in the neutrino spectrum, which might be measured in neutrino telescopes.

\section*{Acknowledgments}

W.A. acknowledges discussions with Stefania Gori and financial support by the University of Cincinnati. 
The work of C.-Y.C. is
supported by NSERC, Canada. Research at the Perimeter Institute is supported in part
by the Government of Canada through NSERC and by the Province of Ontario through
MEDT. The work of B.D. is supported by the DFG grant No. RO 2516/5-1. B.D. also acknowledges partial support from the 
TUM University Foundation Fellowship, the DFG cluster of excellence 
``Origin and Structure of the Universe", and the Munich Institute for Astro- and Particle Physics (MIAPP) during various stages of this work. The work of A.S. is supported in part by the US DOE Contract No. DE-SC 0012704. B.D. and A.S. thank the organizers of WHEPP XIV at IIT Kanpur for the hospitality during an earlier phase of this work. 


\end{document}